\RequirePackage{fix-cm}

\documentclass[runningheads]{svjour3}

\journalname{Bulletin of Mathematical Biology}

\usepackage{mathptmx}
\usepackage[numbers,sort&compress]{natbib}
\usepackage{slashbox}
\usepackage{varioref}
\usepackage{amssymb,amsfonts,amsmath}

\usepackage[nearskip=-3pt,captionskip=4pt,
            listofformat=subsimple,
            labelformat=simple]{subfig}

\usepackage{epstopdf}
\usepackage{epsfig}
\usepackage{graphicx}
\usepackage{pict2e}
\usepackage{ifthen}
\AtBeginDocument{%
\let\orgaddcontentsline\addcontentsline
\renewcommand{\addcontentsline}[3]{
    \ifthenelse{\equal{#1}{toc}}
        {\ifthenelse{\equal{#2}{subsection}}
            {\orgaddcontentsline{#1}{section}{#3}}
            {\orgaddcontentsline{#1}{#2}{#3}}}
        {\orgaddcontentsline{#1}{#2}{#3}}}}

\usepackage[pdfpagelabels,hypertexnames=true,
            plainpages=false,
            naturalnames=false]{hyperref}
\usepackage{sgame}
\usepackage{color}
\definecolor{darkblue}{rgb}{0,0.1,0.5}
\hypersetup{colorlinks,
            linkcolor=darkblue,
            anchorcolor=darkblue,
            citecolor=darkblue}
\usepackage[normalem]{ulem}

\definecolor{darkgreen}{RGB}{0,127,0}

\labelformat{equation}{\textup{(#1)}}
\labelformat{enumi}{\textup{(#1)}}

\numberwithin{equation}{section}

\let\orgautoref\autoref

\renewcommand{\autoref}
        {\def\equationautorefname{Eq.}%
         \def\figureautorefname{Fig.}%
         \def\subfigureautorefname{Fig.}%
         \def\sectionautorefname{Sect.}%
         \def\subsectionautorefname{Sect.}%
         \def\subsubsectionautorefname{Sect.}%
         \def\Itemautorefname{item}%
         \def\tableautorefname{Table}%
         \orgautoref}

\begin{document}

\title{Formation of Dominance Relationships via Strategy Updating in an Asymmetric
Hawk-Dove Game
}

\titlerunning{Dominance Relationships Via Strategy Updating in an Asymmetric Hawk-Dove Game}        


\author{Jasvir K. Grewal \and Cameron L. Hall \and Mason A. Porter \and Marian S. Dawkins}

\authorrunning{J. K. Grewal, C. L. Hall, M. A. Porter, M. S. Dawkins} 

\institute{
			J.~K. Grewal  \at      
			  Oxford Centre for Industrial and Applied Mathematics, Mathematical Institute, University of Oxford \\      
              Tel.: +44-7877-685096\\
              \email{jasvirkaurgrewal@ymail.com}           
           \and
			C.~L. Hall \at
              Oxford Centre for Industrial and Applied Mathematics, Mathematical Institute, University of Oxford \\
              Tel.: +44-1865-280618\\
              Fax: +44-1865-270515\\
              \email{hall@maths.ox.ac.uk}           
           \and
           M.~A. Porter \at
               Oxford Centre for Industrial and Applied Mathematics, Mathematical Institute, University of Oxford \\
               CABDyN Complex Centre, University of Oxford\\
              Tel.: +44-1865-270687\\
              Fax: +44-1865-270515\\
              \email{porterm@maths.ox.ac.uk}           
           \and
           M.~S. Dawkins \at
              Department of Zoology, University of Oxford\\
              Tel.: +44-1865-271234\\
              Fax: +44-1865-310447\\
              \email{marian.dawkins@zoo.ox.ac.uk}           
}

\date{\today}

\maketitle




\begin{abstract}

We develop a model to describe the development of dominance relations between social animals as they use past experiences to inform future interactions. Using the game-theoretic framework of a Hawk-Dove game with asymmetric resource-holding potentials (RHPs), we derive a simple model that describes the social interactions of animals that compete for resources. We then consider a game-playing strategy in which animals acquire information about their RHP asymmetry from the results of their previous contests and subsequently use their asymmetry assessment to inform their behavior in future contests. We examine how directly incorporating the fact that animals have incomplete information in their interactions can lead to outcomes that differ from what would be expected if one considers the situation as a static game in which the animals have perfect information about the asymmetry size. We thereby obtain results that are consistent with observations of dominance-hierarchy formation in social animals. We also discuss how increased interactions between animals can speed up the asymmetry assessment process and how this can explain why aggression can sometimes decrease as the number of interactions between animals increases over time.


\end{abstract}


\section{Introduction}
\label{intro}


\subsection{Background}
\label{Background}

From tiny insects to large mammals, nature includes many examples of animals that live together in groups. Such social animals gain several benefits from living in a group, including the dilution of predation risk and increased foraging success \cite{group}. However, the limited availability of resources (such as water, food, and space) forces social animals to divide restricted resources amongst themselves. This can involve peaceful exchanges, violent fights, or other forms of social interaction; and different groups of social animals have developed different strategies for determining what type of behavior to exhibit when interacting with other animals in their group. This variety in the strategies observed in social animals makes the study of social behavior a rich area for research. It also yields several problems that are amenable to mathematical investigation.

In the evolution of social behavior, a key factor is whether participants meet just once or whether they meet repeatedly \cite{trivers}.  In the famous example of the Prisoner's Dilemma, it is always best for a pair of participants to defect or behave selfishly in a one-off interaction. However, the Prisoner's Dilemma can yield cooperation in the case of repeated interactions \cite{axelrod, nowak3, nowak2, Lehman}: participants can achieve a better outcome if they cooperate, even to the point of giving or sharing valuable resources. The so-called `shadow of the future' (i.e., the future consequences of failing to reciprocate) can imply that giving up or sharing a resource on one occasion is beneficial in the long run because of the many future occasions in which a cooperator receives resources in return \cite{axelrod}.  Such a notion of delayed reciprocity, which arises amidst repeated interactions among a group of individuals, is sometimes called 
`reciprocal altruism.'

The evolution of reciprocal altruism has received a great deal of theoretical attention \cite{dawkinsasymm, Lehman, nowak2}, despite the relative paucity of empirical evidence for it \cite{dugatkin1, cluttenbrock2}.  Meanwhile, there have been comparatively few studies of the evolution of the much more established and rather widespread phenomenon of dominance hierarchies among social animals.

Dominance hierarchies, which also arise from repeated interactions among the same individuals, occur in many groups animals --- including birds, fish, mammals, and insects \cite{Barnard, rushen, Broom2} --- and they are interesting theoretically because animals appear to behave altruistically once such a hierarchy has been established. Subordinates give way to dominants and allow them to take resources without a fight, often without any show of aggression (or even by avoiding any interactions altogether). This raises several key questions: Why should dominance hierarchies arise? Why should subordinate animals stop even attempting to obtain valuable resources? In this paper, we show that dominance hierarchies can arise naturally from a series of animal contests in 
the form of a series of asymmetric Hawk-Dove games that covers situations in which the same participants meet repeatedly.  

The original Hawk-Dove game \cite{smith2} assumes that animals that are in conflict over a resource play either a `Hawk' strategy (i.e., always fight and continue until injured or the opponent retreats) or a `Dove' strategy (i.e., display but retreat if the opponent escalates). In any particular conflict, it is assumed that an animal picks its strategy uniformly at random and also irrespective of previous interactions or any asymmetries such as size differences. A simple modification of the original Hawk-Dove game introduces a third strategy, `Conditional Hawk', which takes 
asymmetries into account: a Conditional Hawk plays the Hawk strategy if an opponent is smaller but plays the Dove strategy if the opponent is larger. It has been demonstrated that the strategy of `Conditional Hawk with assessment' is superior to both Hawk and Dove \cite{smith, smith3}.  This result has to led to considerable interest in how animals might assess and use information about asymmetries --- especially asymmetries in fighting ability or `resource-holding potential' (RHP) --- to switch adaptively between Hawk and Dove strategies \cite{parker1974, smith3, parkerandrubenstein, Hammerstein, 
arnott}. Body size alone can give some indication of RHP, but it can also be cheated (e.g., via raised hair or feathers) and it is not necessarily a reliable indicator of fighting ability, because RHP also varies with health, nutritional status, and other factors that can experience short-term changes. Animals might therefore use the results of actual fighting to discern whether they are stronger than others \cite{dugatkin3, Mesterton-Gibbons, Bonabeau, Lindquist} and to adjust their behavior as their strength changes with age and experience \cite{Fawcett}.

An important development was the realization that assessment signals --- particularly those that are costly to produce --- can be more reliable indicators of RHP than body size alone \cite{parker1974, parkerandrubenstein}. For example, red deer stags signal their RHPs by engaging in roaring matches \cite{cluttonbrock}. A challenger stag begins by roaring at a low rate, the other stag responds by roaring at a higher rate, and so on. Roaring matches tend to occur between stags who are closely matched in body size, and they tend to be exhausting for both stags. Roaring uses the same thoracic muscles as fighting, so an ability to roar at a high rate (and to maintain the peak rate for a long duration) is highly correlated with fighting ability. The roaring match escalates until one stag retreats or the two stags --- unable to decide relative fighting ability based on roaring rate --- proceed to the next phase of their assessment process (the `parallel walk') or to an actual fight. 

Assessment signals also carry a disadvantage: although they give information about fighting ability without the dangers of actually engaging in a fight, this information is gained at a considerable cost \cite{parkerandrubenstein, dawkins2}. For example, a stag cannot provoke the maximum roaring rate that its opponent is capable of achieving unless it roars at a comparable rate. This is exhausting for both the stag and its opponent \cite{payne}. However, given that a stag's body condition and fighting ability can change considerably over the course of a breeding season, it might still be beneficial for a challenger stag to repeatedly pay the costs of assessment. An older stag might be able to defeat a younger one at the beginning of a breeding season but have a much lower ability to do so after several weeks of defending a group of females without much chance to eat \cite{cluttonbrock}.

When the fighting ability of animals is stable over time, repeated interactions between the same animals can take a completely different form. Instead of paying the often considerable costs of assessment signals for each encounter, animals can use the outcome of previous fights with a given opponent to avoid not only the costs of a fight that it is likely to lose but also the costs of giving assessment signals.  In this situation, one expects dominance hierarchies to form, and our aim in this paper is to show how repeated interactions between the same individuals in an asymmetric game can lead to the formation of stable dominance relations (which in turn yield dominance hierarchies).  In our model, animals gain information about their own and an opponent's RHP through fighting in a series of encounters (rather than through assessment signals before or during a single encounter).  This formulation allows our model to incorporate one of the most striking features of dominance hierarchies in real animals: overt and damaging fighting 
as dominance relations in the hierarchy are being formed, which contrasts starkly with the stable and peaceful stage that arises after a hierarchy has been established \cite{Guhl, rushen, pagel, Ivan1, broom}.

For a long-term strategy of repeatedly playing the Dove strategy without continuous assessment to be successful, which is what one observes among subordinate animals in an established dominance hierarchy, the animals need to have already gained very accurate and reliable information about their own RHP and the RHPs of other animals.  The stakes of an encounter between a pair of animals include not merely one resource but all of the resources that might be lost if an animal persistently assumes a subordinate position in the hierarchy that is inappropriate for its true RHP. The value of accurate information about an animal's own and an opponent's RHP is thus so large that it justifies a high cost to obtain it \cite{dawkins2, payne}, even to the point of engaging in a real fight rather than using the less reliable information that can be gained from assessment signals \cite{pagel}.  With this in mind, we have developed a model in which animals gain high-quality information at the cost of having a small number 
of fights initially, but then receive the benefits of long periods without fighting. This is typical of many of the dominance hierarchies that have been observed in nature \cite{broom}.


\subsection{Outline}

The primary aim of the present paper is to construct a model that allows us to explore the acquisition of information from repeated contests and thereby to examine the development of dominance relationships between individual animals from the same group. Our model provides insight into the temporary but high levels of aggression that are observed in many animal groups when they first form.  

We use game theory to study this problem. This provides a convenient framework in which to analyze the different behaviors that animals can exhibit. (See Ref.~\cite{spencer} for a comparative discussion of several possible approaches.) In the next section, we use the language of game theory to formulate the problem of competition for resources among asymmetric individuals. We begin our analysis with a static description of an asymmetric Hawk-Dove game using the concept of an evolutionary stable strategy (ESS), and we use this as motivation to develop a new but tractable dynamic model in Section~\ref{sec:3}. We discuss our results in Section~\ref{sec:4}, and we conclude and suggest possible extensions to our work in Section~\ref{sec:5}.


\section{Setting up the Game-Theoretic Problem}
\label{sec:2}

In this paper, we work within the field of \emph{evolutionary game theory}, which is concerned with strategic interactions between members of evolving populations of organisms (such as animals or plants). Evolutionary game theory is motivated by the expectation that more successful strategies prevail over time relative to strategies that give lower payoffs. Although evolutionary game theory has now been studied for several decades, even the classical theory is still yielding fascinating new insights \cite{zero,hilbe2013}.


\subsection{Maynard Smith's Hawk-Dove game}\label{HDgame}

As described in Section~\ref{Background}, Maynard Smith created the `Hawk-Dove' game to model animal conflicts \cite{smith2}, in which two animals are contesting some resource and can behave in one of two ways: either like a `Dove' or like a `Hawk' \cite{smith}. If both animals behave like Doves, then they each gain an equal share of the resource. If one animal behaves like a Dove but the other behaves like a Hawk, then the animal that plays the Hawk strategy wins the entire resource. A pair of interacting animals is said to be in a `dominance relation' when one animal persistently behaves like a Hawk and the other persistently behaves like a Dove. In other words, a dominance relation exists between two individuals when one individual attacks or threatens to attack the other, and the other individual shows little or no aggression \cite{Ivan1}. Finally, when both animals play the Hawk strategy, there will be a fight. In his original model, Maynard Smith assumed in this situation that each animal has a 50\% chance of injury and a 50\% chance of winning the fight (by injuring its opponent) \cite{smith}. 

We use the Hawk-Dove model as a foundation for our work. The key difference in the problem that we study is that we do not assume that each animal has an equal probability of winning a fight, because we want to account for the natural asymmetries in animal groups.  This results in an asymmetric game.


\subsubsection{The Evolutionary Stable Strategy}

When he developed the Hawk-Dove game, Maynard Smith also introduced the biologically-motivated concept of an `evolutionary stable strategy' (ESS) with which to analyze the game \cite{smith}.  The definition of an ESS is based on the following idea: if all members of a population adopt the ESS, then a mutant strategy cannot successfully invade the population. 

Because the problem that we study involves an asymmetric game, we need to consider a modified definition for an ESS that describes a strategy for each player. We use similar notation to that in Ref.~\cite{smith}.  We consider a contest between two animals, which we label $A$ and $B$, and we define $\pi _i(X,Y)$ to be the payoff to animal $i \in \{A,B\}$ when animal $A$ plays strategy $X$ and animal $B$ plays strategy $Y$.

\vspace{.1 in}

\noindent \underline{\textbf{Definition:}}
A strategy pair $(I^*,J^*)$ is called an \textit{evolutionary stable strategy (ESS)} if  \begin{displaymath} \pi _A(I^*,J)> \pi _A(I,J) \hspace{10pt} \textup{for all} \hspace{10pt} I \neq I^* \end{displaymath} 
\noindent and 
\begin{displaymath} \hspace{6pt} \pi _B(I,J^*)> \pi _B(I,J) \hspace{10pt} \textup{for all} \hspace{10pt} J \neq J^*. \end{displaymath}

\vspace{.2 in}

Note that the existence of an ESS does not guarantee that a population will achieve that ESS. The definition of an ESS requires the majority of a population to be playing that strategy, and such a strategy might be `inaccessible' \cite{nowak}. Nevertheless, the concept of an ESS is useful, as it can provide insight into the possible evolutionary states that a population can obtain.


\subsection{Modelling the Problem}
\label{assumptions}

In this paper, we aim to show how repeated interactions between the same individuals in an asymmetric game can lead to the formation of stable dominance relations. We thus make the assumption that the probability of repeated interactions between group members is high, so that the benefits of being in a hierarchy can be sufficient to warrant the costs that are associated with establishing a set of dominance relations that yield a dominance hierarchy \cite{dawkinsasymm}.

We also follow four of the assumptions\footnote{These assumptions are in addition to any assumptions that automatically arise from formulating the problem that we study as a game.} that Maynard Smith made in Ref.~\cite{smith}: (i) the animals are contesting a resource whose amount is limited; (ii) the resource is divisible; (iii) all contests between animals are pairwise; and (iv) all reproduction is asexual (in the sense that offspring are identical to their parents in strategy choice and RHP). The assumption about reproduction is necessary in order to make the concept of an ESS meaningful: although our work concentrates entirely on the `short' timescale on which a social structure forms within a group of animals --- as opposed to the `long' timescale on which social strategies might evolve across generations --- we implicitly assume that there is some evolutionary pressure that pushes animals to pick a social strategy that maximizes the resources that they can collect (and hence drives the population towards an ESS.)

In Ref.~\cite{smith3}, Maynard Smith and Parker discussed conditional strategies, which involve using visual cues or costly signals to assess asymmetry size before a contest. They also discussed the gaining of information during a contest. However, the formulation in Ref.~\cite{smith3} assumes that an individual uses only the result of the current round to decide whether to continue a contest for a subsequent round. In the present paper, by contrast, we assume that prior information acquired through \emph{all} previous rounds is used to update strategy choices.  We also assume that asymmetries are assessed through actual fighting rather than through costly signals. Additionally, we introduce additional components into the fighting costs to reflect the realistic feature that losers and winners both experience injury, but to different extents.

Importantly, we make the simplifying assumption that the `resource-holding potentials' (RHPs) of the animals in a group do not vary significantly over time. (An RHP measures an animal's capacity to defeat an opponent who is playing the same strategy.)  We make this assumption because we need RHPs to remain constant over a sufficiently long period to ensure that the constant revision of estimates of relative RHP through repeated fighting is not beneficial. We assume that the total amount of the resource is fixed, and we normalize its total amount to be 1.

We want to model how animals interact with each other. As in the classical Hawk-Dove game, we suppose that each animal can play one of two strategies: Hawk or Dove. Although it is more realistic to also consider additional strategies, it is both convenient and reasonable to consider only these two possibilities, because they reflect the main types of behavior that animals exhibit \cite{female, stag, lizard}. 

As we discussed in Section~\ref{HDgame}, three possible results arise from considering a situation in which animals $A$ and $B$ can play the strategies Dove or Hawk: both animals can share the resource, one can withdraw from the contest after the series of displays (so that the resource goes to its opponent), or both animals can escalate the situation into a fight. 

In the final scenario, a fight occurs, and we assume (as in Ref.~\cite{smith}) that each animal will continue fighting until either it is injured and or it is forced to retreat. Thus, two outcomes are possible: either $A$ wins the fight or $B$ wins the fight. Furthermore, fighting includes associated costs that we need to take into acount.  In Ref.~\cite{smith}, Maynard Smith assumed that fighting results in one of the two animals becoming injured and thus forced to retreat. However, it is more realistic to suppose when two animals fight that both the winner and the loser can incur some injury (although to different extents). We let $c_{L} \geq 0$ denote the cost of fighting for the loser, and let $c_{W} \geq 0$ denote the cost of fighting for the winner. We suppose that injuries from a fight include a base cost of being in a fight, which we denote by $c_{0L}$ for the loser and $c_{0W}$ for the winner, as well as an additional cost that depends on the asymmetry in the RHPs of the two animals (i.e., an animal with a lower RHP suffers a higher cost in a fight).  We assume for simplicity that the asymmetry in cost has a linear dependence on the difference in RHP, and we use $c_{1L}$ to denote the additional injury cost for the loser and $c_{1W}$ to denote the additional injury cost for the winner.   In summary, the costs of fighting are
\begin{align*}
	c_{L} & =c_{0L} + c_{1L}\hspace{2pt}y\,,\\
	c_{W} & =c_{0W} - c_{1W}\hspace{2pt}y\,,\\
	c_{0L}\,, \hspace{2pt} & c_{1L}\,, \hspace{2pt} c_{0W} \,, \hspace{2pt} c_{1W}  \hspace{2pt} \geq \hspace{2pt} 0\,, 
\end{align*}
\noindent where the asymmetry in RHP is defined by
\begin{equation}
	y= \textup{RHP of winner} - \textup{RHP of loser}\,.
\end{equation}

We now consider the probability for an animal to win a fight. Once such a probability is determined, it can be used to compute the expected payoff when an animal fights.  We can then summarize the game using a payoff matrix, which in turn allows us to determine the game's ESS. In Ref.~\cite{smith}, Maynard Smith assumed that the two animals have an equal probability of winning a fight. However, we are considering animals with different RHPs, so we also need to incorporate asymmetry into the probability of winning a fight. Suppose that the RHP $s$ of an animal is normalized so that $s \in [0,1]$. We define the difference $z \in [-1,1]$ in the RHPs to be
\begin{equation}
	z = \textup{RHP of animal $A$} - \textup{RHP of animal $B$}\,.
\end{equation}
Using this formulation of asymmetry gives the game shown in Fig.~\ref{fig:2}. Note that the Hawk-Hawk box (in the lower right) includes two possible payoffs for each animal. The entries at the top of the box apply when animal $B$ wins the fight, and the entries at the bottom apply when animal $A$ wins the fight.

\renewcommand{\gamestretch}{3.3}

\def\stackedpayoffs#1#2#3{%
\begin{array}{c}#1\\[1.8mm]#2\\[1.5mm]#3\end{array}
}

\begin{figure}[htb]\hspace*{\fill}%
\begin{game}{2}{2}[Animal~$A$][Animal~$B$]

& Dove & Hawk \\
Dove & $\left(\frac{1}{2},\frac{1}{2}\right)$ &$\left(0,1\right)$\\
Hawk & $\left(1,0\right)$ &$\stackedpayoffs{\left(-c_{0L} + c_{1L} z, 1-c_{0W} - c_{1W} z\right)}{\textup{or}}{\left(1-c_{0W} + c_{1W} z, -c_{0L} - c_{1L} z\right)}$
\end{game}\hspace*{\fill}%
\caption[]{A payoff matrix for incorporating resource-holding potential (RHP) asymmetry into a Hawk-Dove game. The Hawk-Hawk box incorporates the idea that both the winner and the loser in a fight between two animals incur injury (which can be to different extents).}
\label{fig:2}
\end{figure}

We assume that the probability $p_{A_\mathrm{wins}}$ that animal $A$ wins the fight is of the form
\begin{equation}
	p_{A_\mathrm{wins}}= \frac{1}{2}z + \frac{1}{2}\,.
\end{equation}
This choice is analytically tractable, and it is a reasonable simplification because it captures the fact that the probability that animal $A$ wins a fight increases as $z$ increases.  This corresponds to an increase in the RHP of animal $A$ relative to that of animal $B$. Additionally, the two animals have an equal probability of winning the fight when they have the same RHP (i.e., when $z=0$). Because $p_{A_\mathrm{wins}} = 1 - p_{B_\mathrm{wins}}$, it follows that the probability $p_{B_\mathrm{wins}}$ that animal $B$ wins the fight is
\begin{equation}
	p_{B_\mathrm{wins}} = \frac{1}{2} - \frac{1}{2}z\,.
\end{equation}

Using the probability of winning a fight and the costs of fighting (which take an analytically tractable form), we calculate the expected payoff for each animal when there is an escalated contest. Let $\delta_A$ and $\delta_B$ denote the respective expected payoffs for animal $A$ and animal $B$ to fight each other. We assume that these payoffs have the form
\begin{align}
	\delta_A &= \alpha z^2 + \beta z + \gamma\,, \notag \\ 
	\delta_B &= \alpha z^2 - \beta z + \gamma\,,
\end{align}
\noindent where
\begin{align}
	\alpha &= \frac{1}{2} \big(c_{1W} - c_{1L})\,, \notag \\
	\beta &= \frac{1}{2} \big(1- c_{0W} + c_{1W} + c_{0L} + c_{1L})\,, \notag \\
	\gamma &= \frac{1}{2} (1 - c_{0W} - c_{0L})\,.
\end{align}
This yields the game in Fig.~\ref{fig:3}.

\renewcommand{\gamestretch}{3.3}
\begin{figure}[htb]\hspace*{\fill}%
\begin{game}{2}{2}[Animal~$A$][Animal~$B$]
& Dove & Hawk \\
Dove & $\left(\frac{1}{2},\frac{1}{2}\right)$ &$\left(0,1\right)$\\
Hawk & $\left(1,0\right)$ & $\left(\delta_A, \delta_B\right)$
\end{game}\hspace*{\fill}%
\caption[]{The payoff matrix for the expected game between animals $A$ and $B$, where each animal can adopt either a Hawk strategy or a Dove strategy. The parameters $\delta_A$ and $\delta_B$ denote, respectively, the expected payoffs from fighting for animals $A$ and $B$.}
\label{fig:3}
\end{figure}


\subsubsection{Analyzing the Game}
\label{static}

One way to study evolutionary games is to conduct a static analysis, which entails determining the parameter values for which a game has one or more ESSs. In this section, we use such a simple analysis to gain some understanding of the game and to motivate further development of our model. Conveniently, we need not consider the existence of `mixed' ESSs --- where individuals pick their strategy for a particular interaction according to some probabilistic rule --- as it has been proven that asymmetric games do not possess mixed ESSs \cite{selten}.

We use the notation of $X$-$Y$ to indicate that animal $A$ plays strategy $X$ and animal $B$ plays strategy $Y$. By considering the definition of an ESS, one can see that the strategy pairs Dove-Hawk and Hawk-Dove satisfy the ESS conditions for certain values of the parameters that appear in the Hawk-Hawk box in Fig.~\ref{fig:3}. More precisely, these ESSs exist whenever the parameter choices give a positive expected payoff from fighting for one animal and a negative expected payoff from fighting for the other. Whichever animal expects to achieve a negative payoff when there is a fight is the animal that plays Dove in the ESS strategy pair. 

These two ESSs provide some insight into how a dominance hierarchy can be formed in a group of animals. As we discussed earlier, a dominance relation exists between two individuals when one attacks or threatens to attack the other but the other shows little or no aggression. Using the notion of a dominance relation, we interpret the Dove-Hawk ESS as indicating that animal $A$ is dominated by animal $B$, whereas the Hawk-Dove ESS corresponds to animal $A$ dominating animal $B$. Combining all of the dominance relations between members in an animal group yields a dominance hierarchy.  See the discussion in Section \ref{forming}.

When fighting (i.e., the Hawk-Hawk strategy pair) gives a negative expected payoff for each animal, then the Dove-Hawk and Hawk-Dove strategy pairs are both ESSs. This raises an interesting problem of determining which strategy an animal would use in such a situation.  We discuss this issue in further detail in Section~\ref{motivation}.


\subsection{Motivation for Our Model}
\label{motivation}

We seek to examine how the strategies develop over time and how this can lead to a dominance relation. We consider pairwise contests between animals and then use the outcomes of these pairwise interactions to construct a model for the changing behavior of a population of animals over time. 

In our model, animals do not initially know the true size of the asymmetries between themselves and their opponents. We incorporate such uncertainty directly into our model and examine a situation in which each animal estimates the value of its asymmetry with an opponent and then uses the information from this estimate to help determine its behavior. Such estimation is known to occur in nature (see, for example, Ref.~\cite{cluttonbrock}).  Moreover, it clearly makes sense from a modelling perspective to incorporate uncertainty: if animals are certain of differences in RHPs, then the animal with the largest RHP is seemingly very unlikely to share a resource.  However, resource-sharing between members of animal groups has been demonstrated in many different types of species (from honeybees \cite{honeybee} to vervet monkeys \cite{vervet}). For example, vampire bats have been observed to regurgitate blood that they have obtained in order to give it to a hungry member of their colony \cite{vampire}. 

As we discussed above, which ESS occurs depends on the sign of the expected payoffs from fighting. This provides an important motivation to consider how the strategies adopted by animals evolve when faced with negative expected payoffs. In the context of animal behavior, negative payoffs might correspond to situations in which the costs of fighting are high.  For example, combat can be lethal when animals are fighting over a resource that has a significant value \cite{combat}.

When the expected payoffs from fighting are negative for both players, we need to determine the optimal strategy for the animals to choose. One possibility would be for both players to follow the Dove strategy in order to avoid the possibility of injury associated with a fight. However, this suggests an important question: What stops one animal from defecting from the Dove strategy to obtain the entire resource by playing Hawk? Dove-Dove can never be an ESS, and (as discussed in Section \ref{static}) both Hawk-Dove and Dove-Hawk are possible ESSs when the expected payoff from fighting is negative for both players.

This problem is reminiscent of the iterated Prisoner's Dilemma game, in which two players are each given the choice to either cooperate or defect \cite{axelrod}. In a single round of the iterated Prisoner's Dilemma, each player chooses rationally whether to defect, despite the fact that both players gain more when they both cooperate. However, cooperation can occur in the iterated Prisoner's Dilemma, in which the same pair of players play the single-round game repeatedly, because players can use information obtained from previous rounds to help them decide which strategy to play. Because each player knows the strategies that his/her opponent has played previously, a choice to defect is now more harmful than it is in a single-round game because it encourages his/her opponent to defect in future rounds. Consequently, both players are more likely to cooperate, and they each earn a significantly better payoff than would have been the case by using any other strategy over the $N > 1$ rounds of the game. The game can be assumed to 
have an infinite horizon. (See Ref.~\cite{termination} for a discussion of termination rules and their effects on cooperation.)

As we have just discussed, cooperation can exist as an optimal strategy in the iterated Prisoner's Dilemma game even though defection is always optimal in a single-round Prisoner's Dilemma. A notable example of this is the `Tit for Tat' strategy \cite{axelrod}, which (infamously) was the winning strategy in a computer tournament of the iterated Prisoner's Dilemma. An individual that plays Tit for Tat cooperates in round 1 of a game; in round $n > 1$, the player simply copies the strategy that its opponent played in round $n-1$. This example highlights how considering repeated interactions, which is the case for the animals in our example, can lead to different strategy choices than in single-round games. It is worth noting, however, that an important limitation of the iterated Prisoner's Dilemma is its dependence on the assumption of symmetric players \cite{dawkinsasymm}.


\section{The Model}
\label{sec:3}

Recall that we are considering contests between two animals $A$ and $B$, and that we are allowing the strategies that they employ to change from round to round. Specifically, we assume that each animal has a probability of adopting the Hawk strategy and a complementary probability of adopting the Dove strategy, and we let these probabilities evolve over time. Let $p_i^{(n)}$ denote the probability that animal $i$ plays Hawk in the $n^{\mathrm{th}}$ round. We seek to determine a relationship between $p_i^{(n+1)}$ and $p_i ^{(n)}$ that, in some sense, optimizes the resources that are gained by each animal over multiple encounters.

With this in mind, we need to construct an updating rule for the probability that each animal uses the Hawk strategy. We hypothesize that an animal's perception of optimal behavior (assuming perfect information about RHPs) is determined by whether its expected payoff from fighting is positive or negative. From the static analysis, we have already seen that Hawk-Hawk is a unique ESS when the expected payoff from fighting is positive for both individuals. Similarly, if animal $A$ has a positive (respectively, negative) expected payoff from fighting and animal $B$ has a negative (respectively, positive) expected payoff from fighting, then the only ESS is Hawk-Dove (respectively, Dove-Hawk).

However, we encounter some difficulty when both animals expect to receive a negative payoff from fighting. From the perspective of classical evolutionary game theory, both Hawk-Dove and Dove-Hawk are ESSs in this situation, and it has been theorized that animals use signals that do not change the RHP to determine who takes the Hawk role and who takes the Dove role \cite{parkerandrubenstein}. That notwithstanding, there are strong similarities between repeated encounters in which both animals receive an expected negative payoff from fighting and the iterated Prisoner's Dilemma. It is easy to envisage a situation in which a large penalty for fighting leads animals to cooperate because pursuing an aggressive strategy risks retaliation in later rounds. Consequently, with each animal possibly being penalized with negative payoffs in the future, we posit that each of them would prefer to play Dove than risk playing Hawk. One can also motivate such behavior as part of an act of mutual restraint that reduces the 
animals' energy costs and injury risks \cite{warblers}. We therefore assume that the preferred behavior of an individual animal depends only on the positivity of its own expected payoff from fighting (regardless of the expected payoff for its competitor).

We use increments to change an animal's strategy by a step towards its perceived optimal strategy at each stage. Recall that $\delta _i$ denotes animal $i$'s expected payoff for fighting. We compute the probability that animal $i$ plays Hawk by using
\begin{align}\label{model:1}
	p_i^{(n+1)}=p_i^{(n)} + k \times \begin{cases} 0 - p_i^{(n)}\,, \hspace{15pt} \mbox{if} \hspace{5pt} \delta _i < 0\,, \\ 
	1 - p_i^{(n)}\,, \hspace{15pt} \mbox{if} \hspace{5pt}  \delta _i  \geq 0\,, \end{cases} 
\end{align}
where $i\in\{A,B\}$ and the `responsiveness level' of the animals is $k \geq 0$.  An animal's responsiveness level measures how quickly it responds to events of previous rounds, and we assume for simplicity that this value is the same for both animals. If $k$ is too large, then there is a chance for an `unlucky' animal with high RHP to find itself as the subordinate after losing a small number of fights by chance. If $k$ is too small, however, then an animal with low RHP might find itself severely injured because it persists in playing Hawk even when it is clear that the expected payoff from fighting is negative. We see from equation \ref{model:1} that situations in which the expected payoff for fighting is positive will increase the probability of choosing to fight during the next round, whereas situations in which the expected payoff for fighting is negative will decrease this probability.

We now incorporate the idea that an animal does not initially know the size of the asymmetry between itself and its opponent but has to find out by having fights. Because of the high levels of aggression that occur initially when animals are placed into a group (see Section~\ref{intro}), we suppose that each animal starts by playing Hawk with probability $1$.  Based on repeated pairwise contests in the game, animals then update their perceptions of the real size of the asymmetry.  As we have discussed previously, the asymmetry size has a significant effect on the probability for an animal to win a fight. It thus has a significant effect on the expected payoffs from fighting and, in turn, on the strategy to be played.

By updating the perceptions of RHP difference of the animals in the game, we model how the behaviors of the animals evolve during the game.  Because the estimated asymmetry size between a pair of animals plays a significant role in the modelling of strategy choices, it is important to use a good updating rule for the estimates of the asymmetry.  As we discuss in Section \ref{bayes}, we use Bayesian updating.


\subsection{Bayesian Updating}\label{bayes}

Recall that $z$ is the difference in RHPs:
\begin{equation}
	z = \textup{RHP of animal $A$} - \textup{RHP of animal $B$}\,.
\end{equation}
We are considering a situation in which animals $A$ and $B$ each estimate $z$ and then update their estimates after each round of the game. To update these estimates from the outcomes of the game, we use Bayesian updating, so beliefs about the true value of the asymmetry $z$ can be expressed as a probability and are updated as new information becomes available \cite{bayesian}.
 
Bayesian updating involves iterated use of Bayes' rule: we update prior estimates $z$ to obtain a posterior estimate of $z$ that incorporates new information from a fight. In our problem, we update the estimates of the asymmetry by calculating
\begin{equation}
	P( z | \textup{fight outcome})  = \cfrac{P(\textup{fight outcome}| z ) \hspace{5pt} \times \hspace{5pt} P( z )}{P(\textup{fight outcome})}\,,
\end{equation}
which gives a probability distribution for $z$. By determining the peak of the posterior, we determine the most likely value of $z$ and use this as the updated estimate of $z$. There is strong evidence that animals are capable of updating, and there is also some evidence for Bayesian updating in particular \cite{animalbayesian}. 


\subsection{Simulation Examples}

In our simulations, we take the prior probability distributions to be Gaussians that are centered at the initial estimates of the actual asymmetry size $z$ and have a standard deviation $0.1$. We run simulations of the model for $N=200$ interactions between a pair of animals.  The game thus has $N = 200$ rounds.

\begin{figure}[htpb]
\centering 
\includegraphics[scale=0.45]{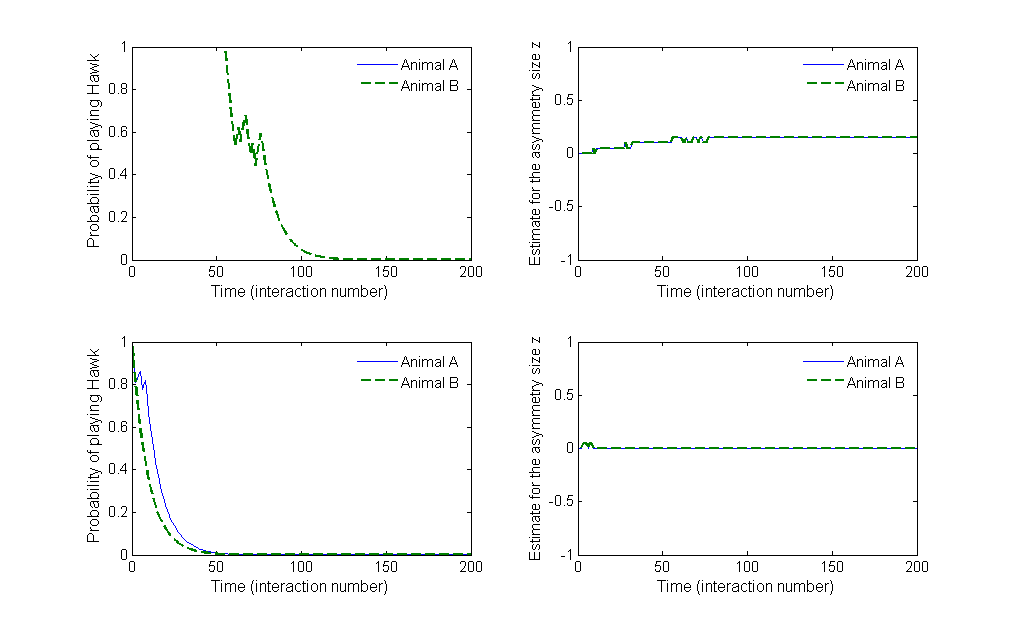}
\caption{Example simulations that illustrate the outcomes of (top) a dominance relation and (bottom) resource-sharing between a pair of animals.  In the left panels, we show the probabilities for animals $A$ and $B$ to play Hawk as a function of discrete time (i.e., the round of the game).  In the right panels, we show the estimates of the size asymmetry as a function of time.
}
\label{fig:4}
\end{figure}

In Fig.~\ref{fig:4}, we show the results of simulations for an actual asymmetry of $z=0.3$ and a responsiveness level of $k=0.1$. In the top panels of Fig.~\ref{fig:4}, we show the results that we obtain by using the parameters values $c_{0W} = 0.2$, $c_{0L}= 0.5$, $c_{1W}= 0.3$, and $c_{1L} = 0.4$.  In the bottom panels, we show the results that we obtain with the parameters values $c_{0W} = 0.4$, $c_{0L}= 0.7$, $c_{1W}= 0.5$, and $c_{1L} = 0.6$.

In the left panels of Fig.~\ref{fig:4}, we show the evolution of the probabilities $p_A$ and $p_B$ as a function of discrete time (i.e., the round of the game).  In the right panels, we show the asymmetry estimates as a function of time. Observe that, in these simulations, the animals have identical estimates in each round. This occurs because each animal starts with the same initial estimate and then updates its estimate at the same rate and using the same information. 

In both situations in Fig.~\ref{fig:4}, the parameters give a positive expected payoff from fighting for animal $A$ but a negative expected payoff from fighting for animal $B$ (i.e., $\delta_A > 0$ and $\delta_B <0$). If we were considering this scenario as a static game in which the animals had perfect knowledge of the asymmetry size $z$ from the start, then we would expect the outcome to be animal $A$ dominating animal $B$ (as this is the ESS). Instead, however, we see that the introduction of uncertainty in the asymmetry size allows other outcomes to occur. For example, the bottom panel of Fig.~\ref{fig:4} shows resource-sharing.

\noindent 

In the top panels of Fig.~\ref{fig:4}, one can see the development of a dominance relation.  Animal $B$ always plays Dove after some transient number of rounds, and Animal $A$ plays Hawk in all 200 rounds. By contrast, resource-sharing behavior develops in the bottom panels.  The difference between the top and bottom panels is that the cost parameters are higher in the latter.  Thus, with the initial conditions held constant, a different cost for fighting can lead to a completely different outcome. In both of the situations shown in Fig.~\ref{fig:4}, the animal strategies reach an equilibrium and become persistent before round $N = 200$ of the game. 

\begin{figure}[htpb]
\centering 
\includegraphics[width = 11cm, height = 5cm]{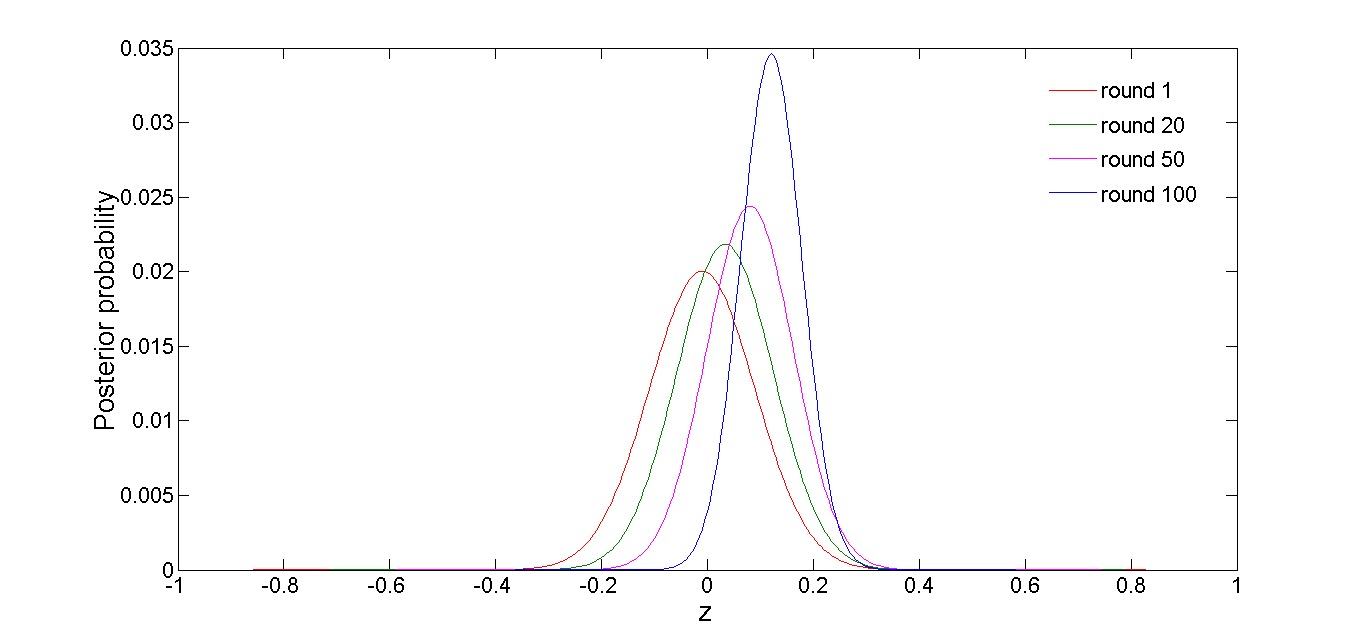}
\caption{An example simulation that illustrates the updating of the posterior distribution for an animal's asymmetry estimate. The initial estimated prior that we use for this simulation is centered around $z=0$, and the `true' value of the asymmetry is $z=0.1$.}
\label{fig:5}
\end{figure}

In Fig.~\ref{fig:5}, we show an example of the updating in time of the posterior-belief distribution for an animal's asymmetry estimate. The parameters that we use to generate this figure are $c_{0W} = 0.2$, $c_{0L}= 0.5$, $c_{1W}= 0.3$, $c_{1L} = 0.4$, $k=0.1$, and $z=0.1$. Recall that the parameter $z$ gives the `true' value of the asymmetry. Observe that the height of the posterior distribution increases gradually as the game progresses.

\noindent 


\section{Discussion}
\label{sec:4}

In this paper, we have developed a model based on the idea that animals acquire information from previous contests with other animals and then use this information to update their strategy choices. For example, when the perceived value of the asymmetry between an animal and its opponent suggests that fighting would provide a negative expected payoff, one expects (and observes in our model) that the animal's behavior becomes less aggressive.

Our model can provide insight into why there is a reduction in aggression as the number of interactions between animals increases over time following high initial levels of fighting when animals are first placed together. Our model also illustrates 
that this phenomenon 
can arise from animals acquiring more information about the value of asymmetries between themselves and the other animals in their group.  As the number of interactions among animals increases, they have more information to use in updating their behavior, so aggression should go down for animals after they learn that being aggressive will cause them more injury than gain.  The key is that increased interactions can speed up the learning process.

\begin{figure}[h]
\centering
\begin{minipage}{.45\textwidth}
\includegraphics[height=55mm, width=50mm]{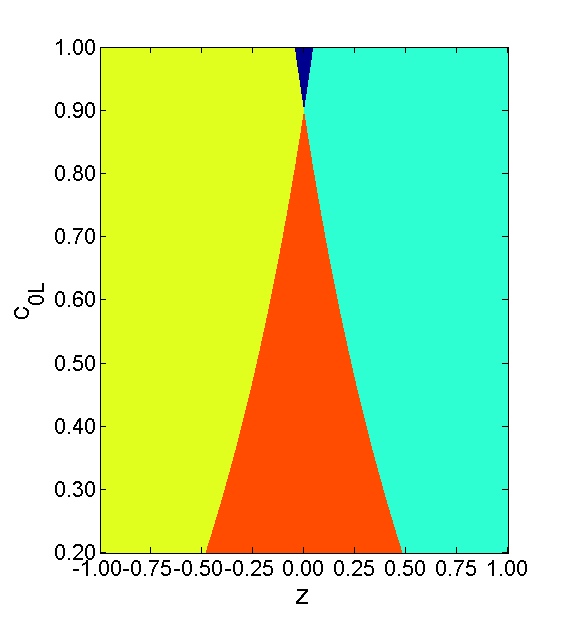}
\end{minipage}
\begin{minipage}{.45\textwidth}
\includegraphics[height=55mm, width=50mm]{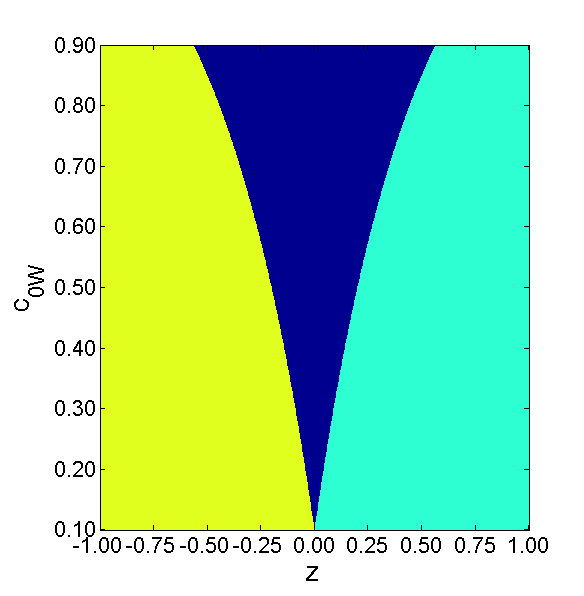}
\end{minipage}
\caption{Bifurcation diagrams illustrating the effects of the size asymmetry $z$ on the relative frequency of resource-sharing versus dominance relations as a function of the cost parameters (left) $c_{0L}$ and (right) $c_{0W}$.  The reddish orange, light greenish blue, greenish yellow, and dark blue regions correspond, respectively, to parameter values that result in persistent Hawk-Hawk, Hawk-Dove, Dove-Hawk, and Dove-Dove behaviors.  (As usual, the left strategy is played by animal $A$, and the right strategy is played by animal $B$.)
}
\label{fig:6}
\end{figure}

In Fig.~\ref{fig:6}, we show that varying the value of the asymmetry size $z$ for fixed cost parameters can lead to different outcomes in our model. In the left panel, we use the parameter values $c_{1L} = 0.2$, $c_{0W} = 0.1$, and $c_{1W}= 0.1$. In the right panel, to illustrate a situation in which there would be more resource-sharing, we use the cost parameters $c_{1L} = 0.6$, $c_{0L} = 0.9$, and $c_{1W}= 0.1$.  Each parameter that we use for the right panel is greater than or equal to the corresponding parameter that we use for the left panel. The reddish orange region corresponds to parameter values that result in persistent Hawk-Hawk behavior. The light greenish blue regions correspond to parameter values that result in persistent Hawk-Dove behavior (i.e., the dominance relation in which $A$ dominates), and the greenish yellow regions correspond to parameter values that result in persistent Dove-Hawk behavior (i.e., the dominance relation in which $B$ dominates). The dark blue regions correspond to parameter values that result in persistent Dove-Dove behavior (i.e., resource-sharing).

In both panels of Fig.~\ref{fig:6}, we observe that there is a window of values for the asymmetry in which resource-sharing can occur (i.e., for which there is no dominance relation). This window increases in size as the cost parameter on the vertical axis increases. From a biological perspective, this corresponds to the fact that a pair of animals with a small difference in their RHPs become more willing to share a resource as the injury costs of fighting become higher.  Given that we have normalized the resource value in formulating our model, an increase in the costs of fighting can alternatively be interpreted as a decrease in the value that the animals attach to the resource.  


\subsection{Constructing a Dominance Hierarchy from a Set of Pairwise Dominance Relations}\label{forming}

As we discussed in Section~\ref{static}, a dominance relation results when a pairwise contest eventually leads either to a Dove-Hawk or to a Hawk-Dove outcome. Resource-sharing results when both animals play the Dove strategy. 

Suppose that there are sequences of pairwise contests among a group of animals.  One can use the eventual outcomes of these contests to construct a dominance hierarchy among these animals. However, there can be difficulties in ranking animals in situations in which the interactions yield dominance relations are not transitive. For example, consider the situation in which animal $A$ dominates animal $B$, animal $A$ shares the resource with animal $C$, but animal $B$ dominates animal $C$. See Ref.~\cite{triads} for examples of intransitive `triads' occurring in animal groups.

A \emph{dominance measure} provides one method with which to rank animals as a dominance hierarchy. The interactions between animals can also be summarized into a \emph{dominance matrix}, which can then be used to rank the dominance relations as a dominance hierarchy \cite{review, dominanceindices, vries}. 

It is potentially desirable to discuss an example of how a model like ours, which produces pairwise dominance relations, can be used as a tool to construct a dominance hierarchy. However, doing so would require us to make several further assumptions, as there are many factors that need to be considered when trying to construct a hierarchy. What structure (if any) should one assume for the interactions between animals? For example, is it suitable to consider a round-robin tournament?  More generally, one can envision (at least in principle) using any time-dependent network of interactions between pairs of animals. Additionally, what dominance measure should one use? 

It is important to note that different ranking methods yield different dominance hierarchies using the same set of dominance relations.  There have been some comparative studies that have examined the relative performance of various ranking methods (see, for example, Ref.~\cite{dominanceindices}).

Other factors also need to be considered when determining a dominance hierarchy. How should one account for situations in which there is a draw in the dominance measure between two or more animals? One way to eliminate draws from is by calculating so-called ``power ratings" for the animals (as is done in some systems for ranking sports teams) \cite{power}. However, from a biological perspective, a draw in a dominance measure might correspond to a situation in which there is no stable dominance relation between a pair of animals that achieved a draw. Consequently, in such a situation, one animal does not rank higher or lower than the other animal; they might well just continue fighting or fight again later.  One can imagine this occurring between a pair of animals when the size of the asymmetry between them is small.


\subsection{Limitations of Our Model}
\label{goodvsbad}

Now that we have developed a simple but reasonable model, a natural desire is to test it using real data. To frame our discussion of this issue, recall that the parameters in our model are the size asymmetry $z$, the animal responsiveness $k$, and the cost parameters $c_{0L}$, $c_{1L}$, $c_{0W}$, and $c_{1W}$.  Our model also includes a probability distribution to model animal uncertainty in estimating $z$, and we can use the same modelling framework with any choice of distribution.

To test our model using empirical data, one would need to determine reasonable estimates for the values for the above parameters (so that we can use them in our simulations). In principle, one should be able to determine values for the parameter $k$ based on research that has been conducted into the responsiveness of many different species of animals to environmental stimuli \cite{responsiveness}. However, it is difficult to measure factors such as the costs of fighting, so it is difficult to test models that contain such parameters against observational data. It is important to note that this limitation (i.e., difficulty in estimating the values of cost parameters using real data) is a weakness that exists throughout the field of evolutionary game theory and is not a limitation that is peculiar to our model.

The evolution of the behavior that animals display depends on the relative sizes of costs and benefits that are associated with adopting different behaviors. This justifies the cost parameters, which are critical for modelling the animals' behavior. One strategy to address the difficulty in measuring such parameters would be to modify our model to use `surrogate' cost parameters such as the duration of an escalated contest \cite{duration} or values that are calculated by considering features such as animal heartbeat rate \cite{heartbeat}. Additionally, some researchers have tried to directly assess the fighting abilities of animals (see, for example, Refs.~\cite{assessment,labra}). 

In our model, we made the assumption that the difference in RHP between each pair of animals remains constant. Our model would need to be generalized to examine groups of animals in which the RHPs can change. For example, the RHPs of red deer stags can change considerably during the breeding season \cite{cluttonbrock}, and this results in frequent serious fighting during the entire season. By contrast, our model is suitable for animal groups, such as flocks of chickens, in which the animals' body condition does not change much over time. Interactions in this type of animal group result in serious fighting initially, but subsequently there is a long period without fighting or assessment.
In such animal groups, our model can provide insight into why this phenomenon occurs --- interactions form part of a learning process in which animals acquire more information about the value of the asymmetries between themselves and the other animals in their group.  This leads to less aggressive behavior in animals who perceive that being aggressive would cause them more injury than gain.  


\subsection{Generalizations of Our Model}
\label{general}

Despite its limitations, the model that we have introduced in this paper has several novel features, and it provides a sound foundation upon which to build.  Indeed, there are many possible ways to generalize our model.  

For example, it is straightforward to incorporate features such as the ability to allow new animals to join a group. In such a scenario, the animals that are already in a group usually show a high level of `in-group' aggression to the new individual \cite{cords, strangers}. From the perspective of our model, such behavior occurs while the animals are acquiring information with which to update their asymmetry estimates. Additionally, our model only considers the possibility of one type of asymmetry between animals. One can, however, incorporate additional asymmetries, such as in their perceived value of a resource.
 
It is also possible to relax some of the assumptions that we made. For example, one could go beyond pairwise contests and consider `bystander effects,' as an animal that observes other animals' contests obtains information by being a spectator. Such effects have been considered in some hierarchy models \cite{dugatkin2}.


\section{Conclusion}
\label{sec:5}

We have developed a novel model, which accounts for asymmetries between individuals, to explain the formation of dominance relations in animals.  In this model, which introduces a connection between Hawk-Dove games and the formation of dominance hierarchies, we examine how directly incorporating the feature that animals initially have incomplete information about their RHP relative to other animals can lead to widespread aggression, followed by the formation of dominance 
relations in animal groups when there might otherwise have been resource-sharing. 

Our model incorporates the fact that it is unlikely for animals to persistently risk injury in situations in which fighting would give a negative expected payoff. The sizes of the asymmetry between animals also play a significant role in determining the outcomes of interactions between members of animal groups. When an animal plays the repeated game (in which it repeatedly fights another animal), it gains information that allows it to update its asymmetry estimate and reduce its aggression level when fighting would cause more injury than gain. This gathering of information from previous contests to update asymmetry estimates can provide insight into biological features in several species of animals, such as the reduced aggression (after initial periods of significant aggression) among a group of animals when there are many repeated interactions. 

Our model provides a solid foundation for a family of models that generalize our study in various ways, and we believe that such models can yield interesting insights on dominance hierarchies and resource-sharing in animals.
 
 
\section*{Acknowledgements}
 
We thank Aaron Clauset for useful comments.


\end{document}